\documentclass[useAMS,usenatbib]{mn2e}
\usepackage{graphicx}
\usepackage{enumerate}

\def\eg{{\it e.g.}}

\def\ie{{\it i.e.}}

\def\pmb#1{\setbox0=\hbox{$#1$}%
  \kern-0.25em\copy0\kern-\wd0
  \kern.05em\copy0\kern-\wd0
  \kern-0.025em\raise.0433em\box0}
\def\ba{\;\pmb{\mit a}}
\def\bx{\;\pmb{\mit x}}
\def\bv{\;\pmb{\mit v}}
\def\bL{\;\pmb{\mit L}}

\long\def\Ignore#1{\relax}

\long\def\strout#1{\setbox0=\hbox{#1}\setbox1 =\hbox to \wd0{\hrulefill}
\hbox{\vbox to \ht0{\box0\kern-3.3ex\hbox{\kern-\wd0\box1}}}}

\def\LaTeX{L\kern-.36em\raise.3ex\hbox{a}\kern-.15em
    T\kern-.1667em\lower.7ex\hbox{E}\kern-.125emX}

\title[Relaxation in simulations]{Relaxation in $N$-body simulations of spherical systems}

\author[J. A. Sellwood]{J. A. Sellwood\thanks{E-mail:
    sellwood@rutgers.edu} \\
 Department of Physics and Astronomy, Rutgers University, \\
    136 Frelinghuysen Road, Piscataway, NJ 08854, US}
\begin{document}
\label{firstpage}

\maketitle

\pagerange{\pageref{firstpage}--\pageref{lastpage}} \pubyear{2015}

\begin{abstract}
I present empirical measurements of the rate of relaxation in $N$-body
simulations of stable spherical systems and distinguish two separate
types of relaxation: energy diffusion that is largely independent of
particle mass, and energy exchange between particles of differing
masses.  While diffusion is generally regarded as a Fokker-Planck
process, it can equivalently be viewed as the consequence of
collective oscillations that are driven by shot noise.  Empirical
diffusion rates scale as $N^{-1}$ in inhomogeneous models, in
agreement with Fokker-Planck predictions, but collective effects cause
relaxation to scale more nearly as $N^{-1/2}$ in the special case of a
uniform sphere.  I use four different methods to compute the
gravitational field, and a 100-fold range in the numbers of particles
in each case.  I find the rate at which energy is exchanged between
particles of differing masses does not depend at all on the force
determination method, but I do find the energy diffusion rate is
marginally lower when a field method is used.  The relaxation rate in
3D is virtually independent of the method used because it is dominated
by distant encounters; any method to estimate the gravitational field
that correctly captures the contributions from distant particles must
also capture their statistical fluctuations and the collective modes
they drive.
\end{abstract} 


\section{Introduction}
The topic of relaxation driven by stellar encounters in star systems
has a long and distinguished history \citep[\eg][]{Chan41, BT08} and
has important implications for the evolution of star clusters
\citep{Spit87} and of active galactic nuclei \citep{Merr13}.
However, it is widely believed that relaxation through
star-star encounters occurs at a negligible rate in the bulk of
galaxies, and therefore $N$-body simulations of galaxies should mimic
this collisionless property.  The study presented here focuses on just
one small aspect of the general problem of relaxation in simulations.

Although the computational power available to simulators has risen
steadily over time, calculations with billions of particles are still
not routinely possible and most simulations employ fewer, more massive
particles.  A rough estimate of the relaxation time $t_{\rm relax}$ in
a simulation of $N$ equal mass particles is \citep[\eg][]{BT08}
\begin{equation}
t_{\rm relax}/t_{\rm cross}\sim 0.1 N/\ln N, 
\end{equation}
with $t_{\rm cross}$ being a typical crossing time.  This, as other
more precise expressions for Fokker-Planck diffusion, contains the
Coulomb logarithm that arises from integration over impact parameters,
implying that every decade of impact parameter makes an equal
contribution to the integral.  It should be noted that the approximate
formula (1) applies for pressure-supported systems, not discs
\citep{Sell13}, and neglects collective effects.  Furthermore, it was
derived for nominally point mass particles, since the argument of the
logarithm comes from the ratio of the system half-mass radius to the
scale on which scattering causes large deflections, whose
contributions would otherwise be overestimated.  It therefore reflects
spatial resolution, which is usually determined by other factors, such
as particle softening or grid resolution in $N$-body codes.
Therefore, $t_{\rm relax}$ should be simply proportional to $N$ when
resolution is held fixed.

Numerical methods used to compute the gravitational field in
simulations that aspire to be collisionless fall into three broad
categories.  The most popular direct method is some type of tree code
\citep[\eg][]{BH86, Spri05} that effectively sums the attraction of
every particle pair, with a softening kernel to limit the magnitude of
the acceleration at short range.  The far more efficient
\citep{Sell14} particle-mesh methods \citep[PM, \eg][]{HE81}
determine the gravitational field on a raster of points that has some
appropriate geometry; forces at the actual particle positions are
computed by interpolation between grid points.  Finally, the least
popular are field methods \citep[\eg][]{CB72, HO92} that expand the
density and potential in a basis set of functions that should be
chosen such that truncating the expansion at low order yields an
adequate approximation to the field.

Weinberg, in a series of papers \citep{Wein99, HBWK05, WK07a, WK07b}
has argued that field methods are inherently superior to other
$N$-body methods in their ability to hide the lumpiness of the
potential from a set of point masses.  In particular, \citet{WK07b}
assert that field methods have ``relaxation times ...\ orders of
magnitude longer'' than in tree codes.  This claim was based on a
lengthy calculation using Hamiltonian mechanics that I review below.

Since a well-chosen basis can yield an adequate approximation to the
total field from a small number of terms, it may seem reasonable to
expect the resulting potential to be smoother than that computed by
other methods.  However, this argument is beguiling for the following
reasons.  Only the lowest order monopole term has a large value about
which shot noise fluctuations are small, while the amplitudes of the
aspherical terms, which are oscillatory, depend on the almost-complete
cancellation of the $N$ contributions, and are entirely noise-driven
in a spherical model, for example.  It is also true that the number of
values that define the potential in PM codes is $n_{\rm grid}$ and,
since each mesh point typically hosts $\sim N / n_{\rm grid}$
particles, each separate value will be subject to a greater degree of
shot noise.  In direct methods, there are $N$ separate contributions
to the field.  But note that the potential is the double integral of
the density; in effect, the potential kernel, which is monotonic and
has infinite range, implies the field at each point is the sum of
contributions from all the sources.  Thus potential variations are far
smoother than could be supported by an arbitrary function defined by
the same number of values.

\citet[][hereafter HB90]{HB90} used a King model for an experimental
comparison between the relaxation rates in simulations when the
gravitational field was determined by three different methods, and
\citet{HO92} extended their results to include a field method.  They
found that the rate of energy diffusion of particles was only mildly
affected by the method used.  While their evidence was quite strong,
they generally employed only 4096 particles and gravity softening
spoiled the equilibrium of some of their initial models.
\citet{Sell08} showed that grid and field methods performed equally
well in the specific problem of bar-halo friction.

Here I present a more general study of fully self-consistent
equilibrium spheres that uses two distinct measures of the
``relaxation'' rate: the energy diffusion rate reported by HB90, and a
measure of the rate of energy exchange between particle species of
differing masses.  The former measure includes all sources of
relaxation, especially collective effects, while the latter is a more
direct consequence of two-body encounters.

\section{Models}
In order to illustrate the importance of collective effects, I here
report measurements in three different mass models.  All three are
spheres with ergodic (isotropic) distribution functions (DFs) that are
therefore stable equilibria.

The mass models are:

\begin{enumerate}[(a)]
\item {\it Hernquist model} \citet{Hern90} developed a simple
  spherical model having the centrally cusped density profile
\begin{equation}
\rho(r) = {Ma \over 2\pi r(r+a)^3},
\label{eq.Hdens}
\end{equation}
which has the potential $\Phi(r) = -GM /(a + r)$.  Here $a$ is a
length scale and $M$ the finite total mass integrated to infinity.
Hernquist also gave the equilibrium isotropic DF for this density and
potential.  I truncate this model so that no particle has sufficient
energy to stray beyond $r=10a$, which causes the density profile to
taper smoothly to zero at that radius, and discards $\sim 26$\% of the
mass, but leaves the density profile as given by eq.~(\ref{eq.Hdens})
over the range $0 \leq r/a \la 5$.

\item {\it Plummer sphere} \citet{Plum11} introduced one of the most
  widely used spherical mass models in astronomy.  It has the cored
  density profile
\begin{equation}
\rho(r) = {3M \over 4\pi a^3}(1 + x^2)^{-5/2},
\end{equation}
where $x = r / a$ and $M$ is the total mass.  The potential is
$\Phi(r) = -(GM/a)(1 + x^2)^{-1/2}$, and the isotropic DF is that of a
polytrope of index 5 (see BT08).  Applying an energy truncation so
that no particle passes outside $r=10a$ discards only $\sim 3.4$\% of
the mass in this model with its lower density envelope.

\item {\it Uniform sphere} with density $\rho_0 = 3M/(4\pi a^3)$,
  where $a$ is the outer radius.  \citet{PS79} give the isotropic DF
  for a homogeneous sphere with a sharp outer boundary.  The harmonic
  potential in the interior of this unusual model implies that all
  particles have the same orbital frequencies; this model, therefore,
  affords a dramatic illustration of the role of collective effects.

\end{enumerate}

In all cases, I employ two equally numerous sets of particles drawn
from the DF: the masses of particles, $m_i = \mu_i m_*$ are such that
those of one sample have $\mu_i=9$ and the other have $\mu_i=1$, and
$m_*$ is chosen such that the combined density profile is that given
by the above expressions.  \citet[][appendix A]{DS00}
describe an optimal method of drawing particle coordinates from a DF
in such a way as to reduce shot noise in the distribution of energies.
\citet{Sell14} reports results that show the material benefit of this
strategy.

The models are evolved for 100 dynamical times, where $t_0 =
(a^3/GM)^{1/2}$, with a timestep of $0.02t_0$ for the uniform sphere
and Plummer models.  The basic step for the Hernquist model is $0.0125$
but time steps are increased, in this case only, by four successive
factors of 2 at appropriate radii.  I save the instantaneous energy of
a representative set of particles ($2 \times 10^4$ or all, whichever
is the less) after every dynamical time.

\section{Methods}
I employ four different numerical methods to determine the
gravitational field from the particles:

\begin{enumerate}[(i)]
\item {\bf BHT} A tree code that uses the scheme first proposed by
  \citet{BH86}.  I include dipole terms and particle groupings are
  opened when they subtend an angle $> \theta_{\rm max}= 0.5$ radians.
  Forces are softened at short range only using the kernel advocated
  by \citet{Mona92}, with $\epsilon=0.05a$ for the uniform sphere and
  $\epsilon=0.1a$ for the other two models.

\item {\bf S3D} A spherical grid, which is a hybrid PM expansion
  method that expands the non-spherically symmetric part in surface
  harmonics on a set of spherical shells \citep{McGl84, Sell03}.
  Force discontinuities, which arise when particle radii cross, are
  eliminated by adopting linear interpolation between radial shells.
  I tabulate the expansion coefficients at 100 logarithmically spaced
  radii, and expand in azimuth up to $l_{\rm max}=8$.

\item {\bf SFP} A field method that employs a biorthonormal set of
  basis functions \citep{CB72, HO92}, which I name SFP (for smooth
  field-particle) but is also known as SCF.  I expand in azimuth up to
  $l_{\rm max}=8$ and employ radial functions up to $n_{\rm max}=10$.
  I use this method for the Plummer and Hernquist models only.

\item {\bf C3D} I do not use the SFP method for the uniform sphere, but
  employ a PM method that uses 3D Cartesian grid \citep{Jame77}, and I
  set $a = 50$ mesh spaces.  The grid has $129^3$ points, except for
  the lowest $N$ case where it was $257^3$ points in order to allow
  plenty of room for expansion of the particle distribution.  Linear
  interpolation results in the inter-particle force given in
  \citet[][appendix]{SM94}, which is well approximated by cubic
  density kernel with $\epsilon \approx 1.8$ grid spaces
  \citep{Sell14}.

\end{enumerate}
More details of all these methods are given in the on-line manual
\citep{Sell14}.

The choices of numerical parameters in each case are somewhat
arbitrary, but values are typical of those used in practice and are
not varied as the particle number is changed.  The rate of relaxation
will depend only weakly on spatial resolution, since only short range
scattering is affected by changes to the softening length, for
example, leading to a slight change in the value of the Coulomb
logarithm.

Truncating the expansion at low order in field methods smooths the
mass distribution, and more aggressive truncation will give rise to a
smoother potential, which must reduce the relaxation rate; \eg\ in the
extreme case of a single term, the particles will be moving in an
almost fixed potential, and no evolution and little relaxation could
occur.  Since the purpose of simulations is to compute the
self-consistent evolution as the density changes, I include sufficient
terms to be able to follow changes that might be expected in an
evolving model.

The tree code uses explicit particle softening, while short-range
forces are implicitly smoothed in the Cartesian grid.  These methods
therefore do not yield the exact Newtonian potential of the mass
distribution.  In order to ensure that the initial model is in
equilibrium in the tree code, I use the largest $N$ simulation of each
type to tabulate the difference between the spherically averaged
central attraction of the particles at $t=0$ and the analytic central
attraction at a 1D array of points; I then interpolate from this table
a supplementary central attraction that is added to the
tree-determined force on each particle before its motion is advanced.
I use a similar procedure for the 3D Cartesian grid, but in order to
avoid shot noise in the (very small) permanent part of central
attraction, the numerical force is computed from a smooth density
distribution assigned to the grid.  These generally small corrections
are not needed for S3D or SFP methods.

\begin{figure*}
\begin{center}
\hbox to \hsize{
\includegraphics[width=.3\hsize]{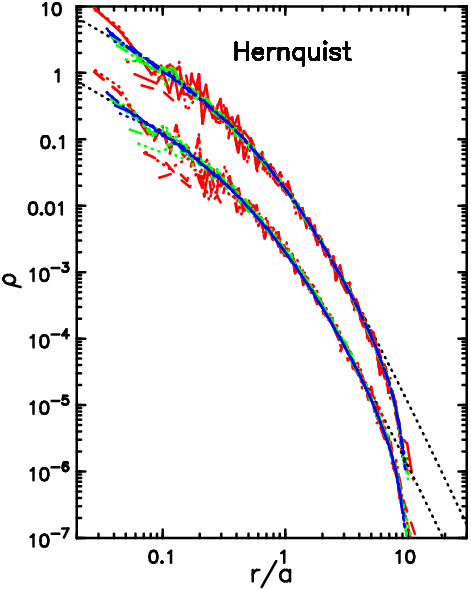}\hfil
\includegraphics[width=.3\hsize]{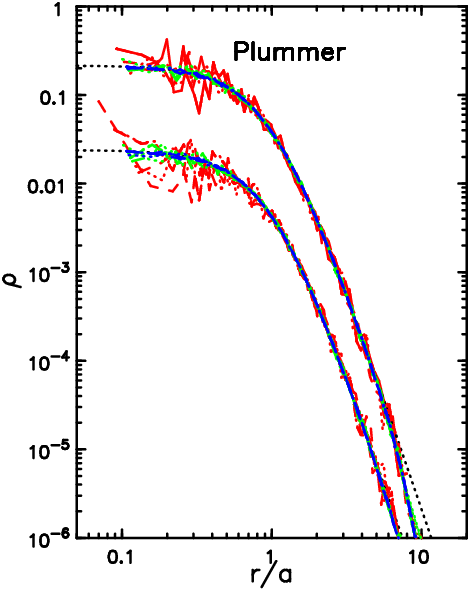}\hfil
\includegraphics[width=.3\hsize]{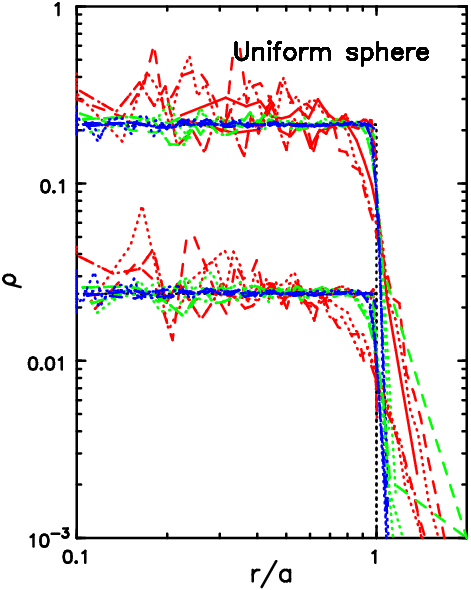}}
\end{center}
\caption{The final density profiles (coloured lines) measured from the
  different particle species that, in all 27 simulations, contribute
  10\% and 90\% of the total density as indicated by the two groups of
  lines.  In each panel, the colour indicates the number of particles:
  red is for $N = 4\times 10^3$, green is for $N = 4\times 10^4$ and
  blue is for $N = 4\times 10^5$.  The different methods for computing
  the field are distinguished by the line styles: solid is for S3D,
  dotted is for BHT, and dashed is for either SFP (Hernquist and
  Plummer) or C3D (uniform sphere).  The black dotted curves indicate
  the expected density profile of each mass species; no particles in
  the left and middle panels started with enough energy to pass beyond
  $r=10a$.}
\label{fig.rho}
\end{figure*}

With these fixed correction terms in C3D and BHT, and with a
coordinate centre for the S3D and SFP methods, it is important to
ensure that the initial set of particles is at rest and centred in
this coordinate frame.  After creating the initial set of particles, I
therefore adjust the positions and speeds by a small amount to ensure
the centre of mass is at the coordinate centre and the model has no
net momentum.  An alternative strategy that would achieve the same
outcome would be to shift the grid (or expansion) centre at frequent
intervals.  I also experimented with inserting mirror pairs of
particles having coordinates $(\bx,\bv)$ and $(-\bx,-\bv)$, in a step
towards a quiet start, but this strategy had the undesirable (for this
study) effects of eliminating any lop-sided, and emphasizing the
bi-symmetric, contributions to the total field.

The instantaneous energy of the $i$th particle is approximately
\begin{equation}
{\cal E}_i = m_iE_i(t) = m_i\left[\Phi(\bx_i) + \textstyle{1\over2}v_i^2\right],
\label{eq.Edef}
\end{equation}
where $E_i$ is the specific energy of the particle, or energy per unit
mass, $\Phi$ is the estimated gravitational potential at the particle
position, $\bx_i(t)$, and $v_i(t)$ is the scalar speed of the
particle.  This definition is exact only for a particle of
infinitesimal mass, since a finite mass particle contributes to
$\Phi$.  The energy required to disperse a system of gravitating
particles, the total energy, is ${\cal E}_{\rm tot} = T + W$, where
$T=\sum_i {1\over2} m_iv_i^2$ and $W = {1\over2} \sum_i m_i
\Phi(\bx_i)$, and the $W$ term is halved because the summation over
the $\Phi$ values includes every pair of particles twice.  The total
energy, defined this way is very well conserved in the simulations,
whereas the sum of the energies (eq.~\ref{eq.Edef}), $\sum_i{\cal E}_i
= T + 2W$, is clearly {\em not\/} the total energy, and is not
conserved.

In order to test for virial equilibrium of an $N$-body simulation, it
is better to measure the virial of Clausius, $W_c \equiv \sum_i m_i
\bx_i\cdot\ba_i$, since the inter-particle forces are not perfectly
Newtonian, especially at short range.  It is easy to show that $W_c =
W$ for precisely inverse-square law accelerations.  The particle
distribution is in equilibrium when $2T=|W_c|$.

It is convenient to chose units such that $G=M=a=1$, so that the
dynamical time $t_0=(a^3/GM)^{1/2}=1$, for example.  A convenient
scaling to physical units for the inhomogeneous models is to choose $a
= 3\;$kpc and $t_0 = 10\;$Myr, which implies $M = 6 \times
10^{10}\;$M$_\odot$ and velocities scale as $(GM/a)^{1/2} \simeq
293\;$km/s.

\section{Results}
Figure~\ref{fig.rho} shows the final ($t=100t_0$) density profiles in
all 27 simulations.  The three panels show the different mass models.
Within each panel the curves drawn in red are from the lowest $N = 4
\times 10^3$, those in green employ $4 \times 10^4$ particles, and
those in blue $4 \times 10^5$, with $N/2$ in each mass species so that
the initial densities from each species differ by factors of nine, as
shown.  The field is determined on the S3D grid for the full-drawn
curves, the BHT method for the dotted curves, while the dashed curves
are from either the SFP, for the Hernquist and Plummer models, or C3D
for the uniform sphere.

\begin{figure*}
\begin{center}
\hbox to \hsize{
\includegraphics[width=.3\hsize]{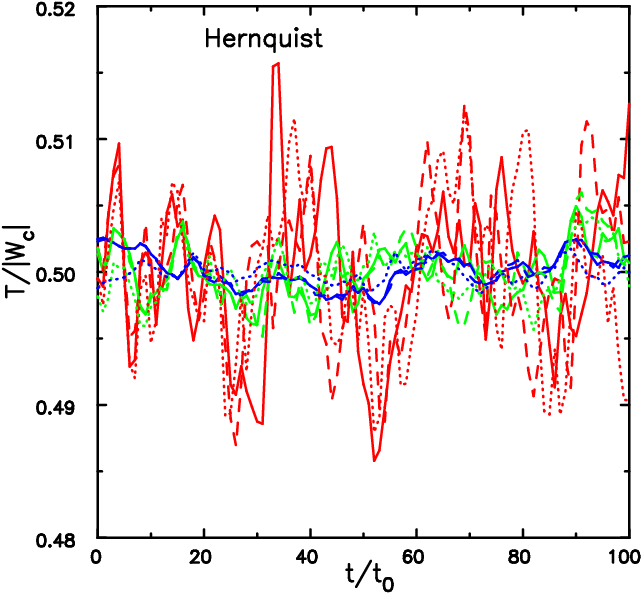}\hfil
\includegraphics[width=.3\hsize]{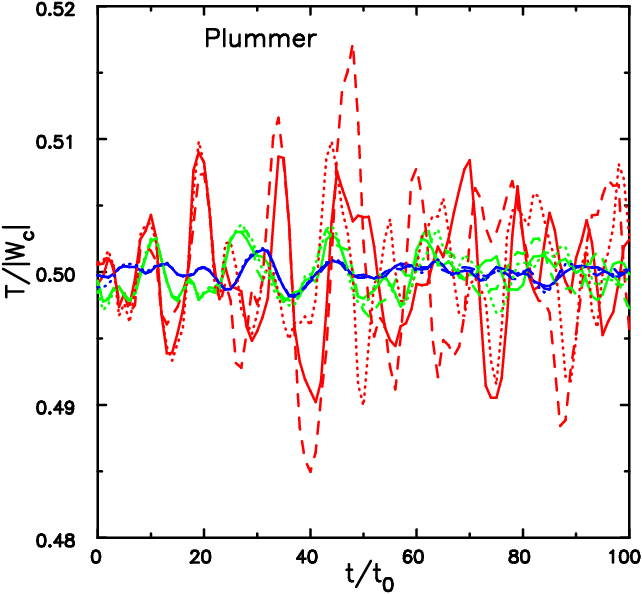}\hfil
\includegraphics[width=.3\hsize]{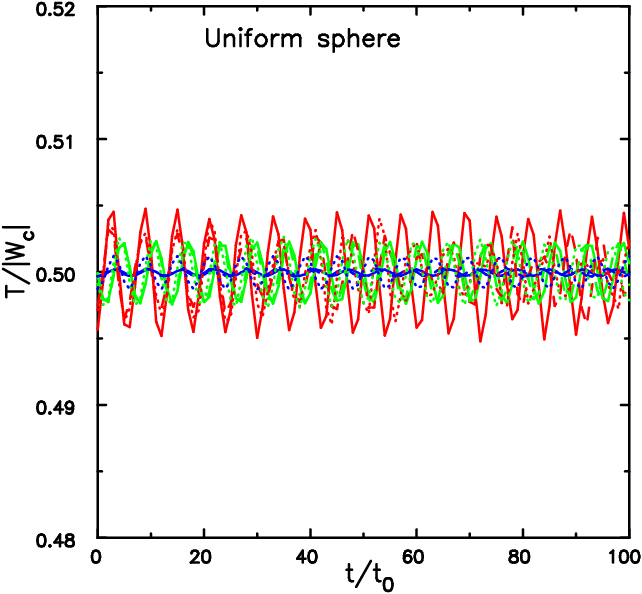}}
\end{center}
\caption{The time evolution of the virial ratio in all 27 simulations.
  As for Fig.~\ref{fig.rho}, the colour indicates the number of
  particles while the different methods for computing the field are
  distinguished by the line styles.}
\label{fig.vir}
\end{figure*}

The density profiles of the models evolve insignificantly for the
largest $N$ (blue curves), confirming that the models are stable
equilibria.  As expected, relaxation drives the greatest changes in
the simulations with the smallest $N$ (red curves) where, in a few
cases, there are hints of some slight segregation of the particles of
different masses over the time interval computed.  \Ignore{Despite the
  variations in the virial ratio, shown in Fig.~\ref{fig.vir},}

\subsection{Collective modes}
Figure~\ref{fig.vir} shows the time evolution of the virial ratio,
$T/|W_c|$ in all 27 simulations reported here.  The three panels show
the different mass models and the colours and line styles are used to
distinguish the particle number and field determination method, as in
Fig.~\ref{fig.rho}.

It is clear from Figure~\ref{fig.vir} that the virial ratio $T/|W_c|$
remains close to $1\over2$ for the duration of all these simulations,
and the fluctuations around this ratio are largest for the lowest $N$
(red curves) and smallest for the highest $N$ (blue curves).  The
fluctuations are aperiodic for the inhomogeneous models, but are
periodic in the uniform sphere, where they have very nearly the same
period, $2\pi t_0$, in all nine simulations.  Furthermore, the initial
behaviour of the curves for models with the same $N$ (colour) is very
similar for the different field determination methods, because the
models were set-up using particles having the same positions and
velocities.  The red curves diverge quite noticeably, but the blue
curves for the SCF and S3D methods remain barely distinguishable to
the end; a slightly different value of $W_c$ arises in the BHT code
because the forces differ due to softening and corrective terms, which
is the reason the dotted curves remain distinct, particularly in the
Hernquist model.

It is helpful to think of these fluctuations as the consequences of
collective modes that are driven by shot-noise in the finite number of
particles, as was recognized long ago by \citet{RR60} in the context
of collisionless plasmas, and has been studied in gravitating systems
by \citet{Wein98} and others.  The simplifying concept here is to
distinguish the ideal collisionless system, which has a set of neutral
and/or damped modes, from the noise spectrum arising from the
particles that excites the modes.  In principle, this approach would
enable the evolution to be calculated by perturbation theory.
Naturally, such an analysis should not differ in its predictions from
the direct evolution of an $N$-body system composed of equal mass
particles, which is exactly what the simulations compute.  The results
presented below reveal that relaxation is largely independent of
particle mass, indicating that global modes are the dominant
relaxation mechanism even when not every particle has the same mass.

The amplitudes of the modes should scale with the shot noise, and the
rms scatter in $T/|W_c|$ about the value 0.5 indeed scales
approximately as $N^{-1/2}$.  The modes have essentially the same
frequency in the uniform sphere where they appear to be almost
undamped.  Modes in collisionless systems can be damped only through
resonant exchanges between the mode and the particles, and conditions
for resonant damping are highly unfavorable because all particles have
the same frequencies in this harmonic potential.

The aperiodic fluctuations in the inhomogeneous models (left and
middle panels) reflect the broader range of frequencies among the
particles in these models.  Collective oscillations that are excited
by shot noise in these cases are almost certainly quickly damped at
resonances, yet the amplitudes fluctuate greatly with neither a clear
decaying, nor a growing, trend.

\begin{figure}
\includegraphics[width=.67\hsize,angle=270]{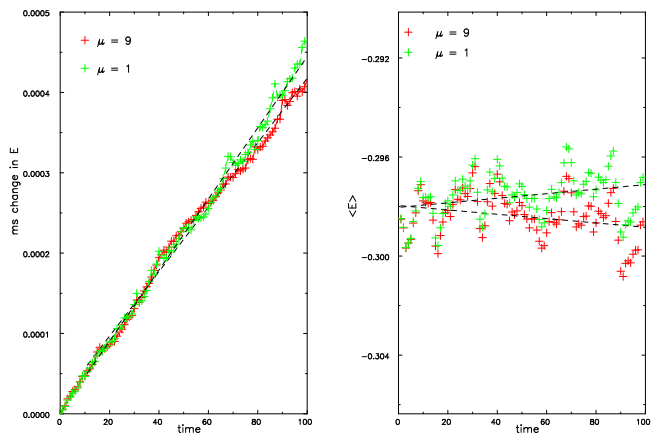}\hfil
\caption{Results from a run of a Hernquist model with $N=2 \times
  10^4$ particles of each mass when forces are determined using the
  field method (SFP).  The left panel shows the mean square change of
  energy of the particles, while the right panel shows the evolution
  of the mean specific energy (per unit mass) of the particles of each
  mass species.}
\label{fig.typical}
\end{figure}

\subsection{Energy diffusion}
The potential in a hypothetical simulation with infinitely many
particles would be smooth and steady, assuming the model is a stable
equilibrium, and the specific energy $E$ of each particle would be
conserved.  Therefore the time evolution of the rms changes in $E$ is
one convenient measure of relaxation in a system with finite $N$.

Figure~\ref{fig.typical} presents a typical set of energy measurements
from the particles in a simulation; values for heavy particles are
drawn in red, while green is used for the light particles.  The left
panel gives the value of $\langle [E_i(t)- E_i(0)]^2\rangle$, the mean
square change since the start in the measured specific energy of the
particles.  It can be seen that the value of this quantity rises
roughly linearly with time, as HB90 found, indicating that the values
are changing through a diffusive process.  I fit a straight line to
the last 90 values (\ie\ ignoring the first 11, where the rise is
often a little steeper), and henceforth report only this fitted slope
as the mean square change of $E$ per dynamical time $t_0$, and its
associated uncertainty.  

The changes presented in Figure~\ref{fig.typical} are averages over
all the particles.  Naturally, the rms changes are larger for those
particles whose orbit periods are shorter.  As this trend is quite
gradual in the inhomogeneous models and non-existent in the uniform
models, the global averages I report in all cases are representative
of those at intermediate radii where the bulk of the particles are
located.

The total angular momentum, $L = |\bL|$, of each particle is also
conserved in a smooth potential, and its changes afford another
measure of relaxation.  I have found that the mean-squared changes in
$L$ also rise roughly linearly with time, but not quite as smoothly as
the $E$ changes in the left panel of Figure~\ref{fig.typical}.
Furthermore, mass segregation is still less evident in the angular
momentum changes.  I therefore focus on the $E$ changes for the
remainder of the paper.

\subsection{Mass segregation}
The right panel of Figure~\ref{fig.typical} gives the time evolution
of the mean specific energy $\langle E_i(t)\rangle$ of the particles
of the separate masses.\footnote{The mean energy measured from the
  simulation, $\langle E_i(0)\rangle \approx -0.30$, is somewhat
  higher than that expected from the DF, $\langle
  E\rangle \approx -0.356$, because the potential well in the
  simulation determined from the selected particles, which are only
  74\% of the total mass, is not as deep as the analytic potential of
  the untruncated model.}  The rapid variations have the same sign for
each particle species because they arise from variations in $T+2W$, as
discussed in \S3, that are related to the virial fluctuations
(Fig.~\ref{fig.vir}).  They result from evolving potential changes
seeded by shot-noise driven fluctuations among the particles.

In addition to these fluctuations, the right panel of
Figure~\ref{fig.typical} displays a slow divergence of the mean
energies of the heavy and light particles, which represents the
gradual exchange of energy that would in the long-run cause the heavy
particles to settle to the centre and the lighter to populate the
envelope of the model.  The slopes of the straight lines fitted to
these data should differ in magnitude by the ratio of the particle
masses, since total energy conservation requires the energy lost by
the heavy particles to be taken up by the light.  In practice, this
symmetry is imperfect because of the large measurement uncertainties.
Henceforth, I report only the slopes, multiplied by $\mu$, and their
statistical uncertainties.

In systems of gravitating particles, heavy particles lose energy to
light particles through dynamical friction \citep{BT08}.  Physically,
a particle is braked by the attraction from its wake -- \ie\ the
response of the surrounding sea of particles to its motion.  In a
system of equal mass particles, the frictional drag on each particle
is balanced, on average, by the scattering accelerations
\citep{Heno73} and no secular changes occur.  But differences between
the braking and acceleration terms cause mass segregation when
particle masses are unequal.

Were energy exchange between particles of the different mass species
the only source of relaxation, the trends in the right panel of
Figure~\ref{fig.typical} would not be strongly masked by short term
oscillations.  That short-term changes are larger than the gradual
diverging trend is evidence that relaxation, measured in the left-hand
panel, is being driven mostly by the collective oscillations of the
model discussed above that are independent of particle masses.

\begin{figure*}
\begin{center}
\includegraphics[width=.38\hsize,angle=270]{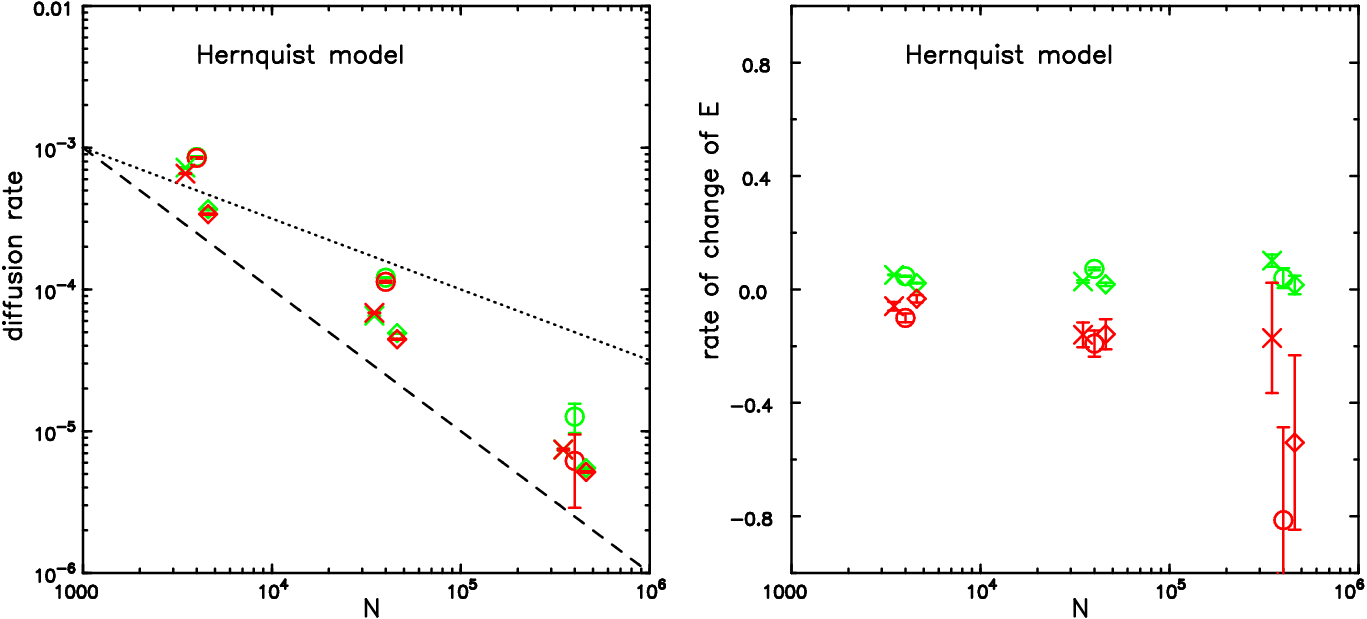}\hfil
\includegraphics[width=.38\hsize,angle=270]{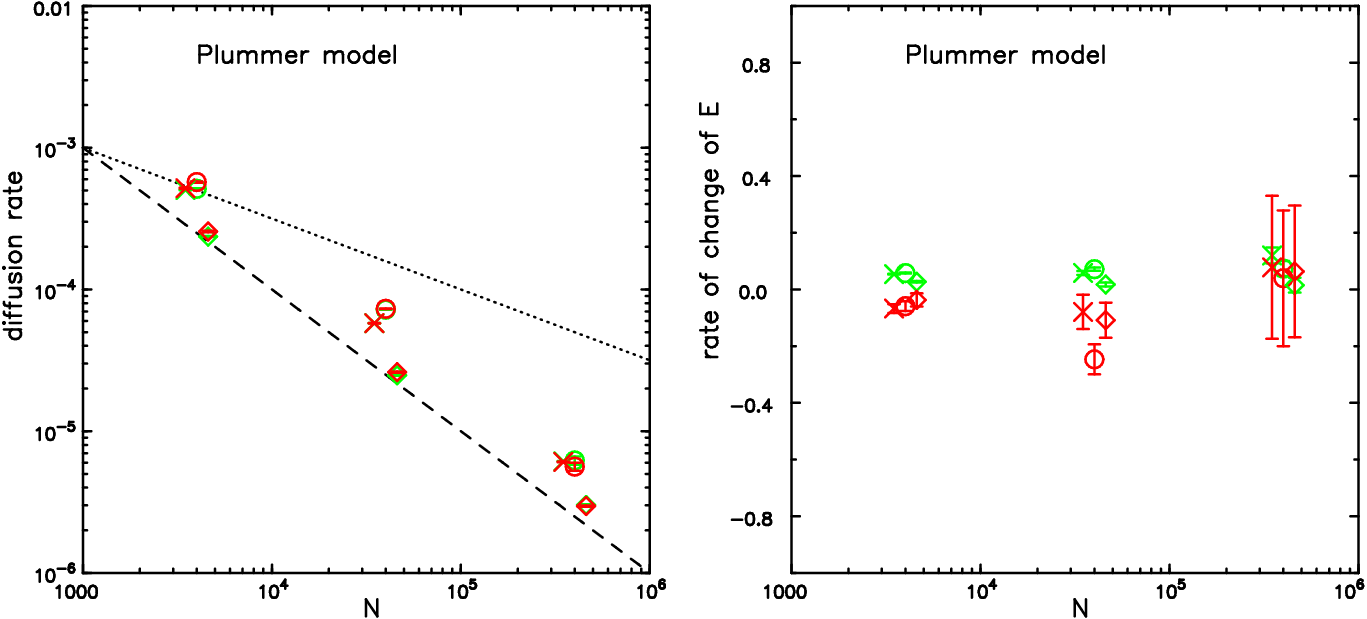}\hfil
\includegraphics[width=.38\hsize,angle=270]{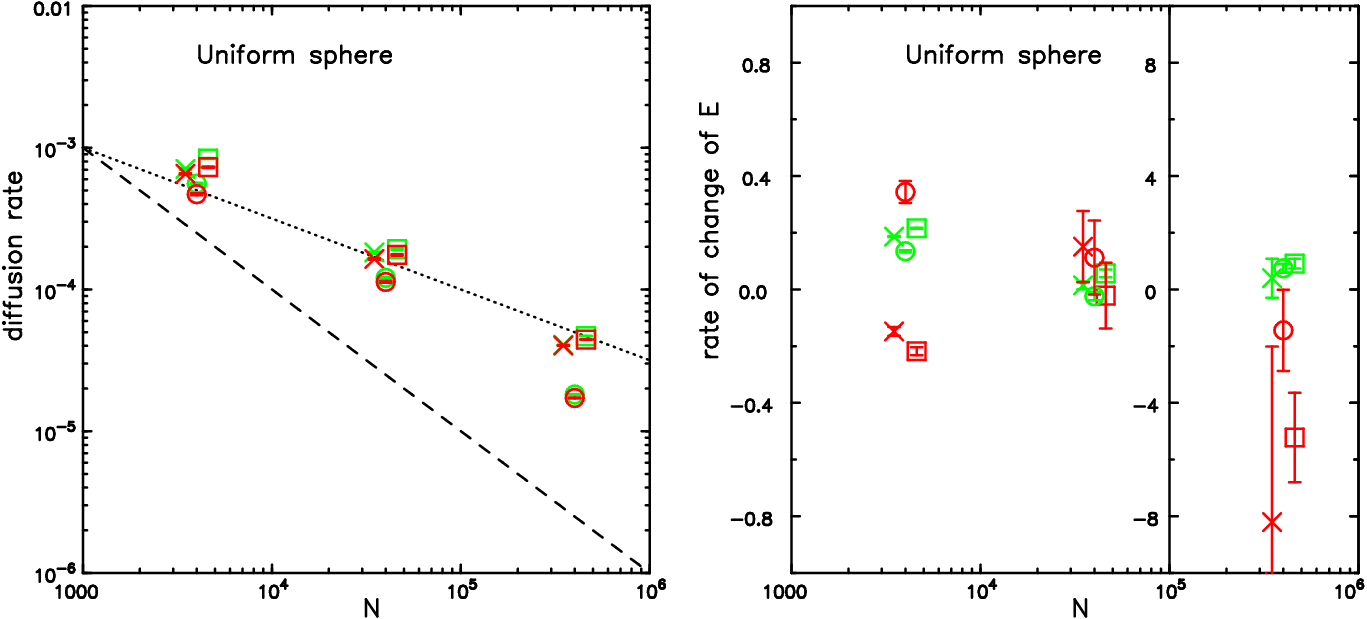}
\end{center}
\caption{Summary of results from 27 simulations.  The left column
  shows the $N$-dependence of the diffusion rate defined in
  eq.~(\ref{eq.edif}) and the right column shows $Nd\langle
  E_i(t)\rangle/dt$, with green points for the light particles and red
  for the nine times heavier particles.  The top row is for the
  Hernquist model, the middle row for the Plummer model and the bottom
  row is for the uniform sphere.  Crosses are from the tree code
  (BHT), circles from the spherical grid (S3D), diamonds from the
  field method (SFP) which was used for the Hernquist and Plummer
  models only, while squares are for a cubic Cartesian grid (C3D)
  which was used for the uniform sphere only.  The error bars show
  $\pm\sigma$ uncertainties in the slopes, and the symbols from the
  different methods have been slightly shifted horizontally for
  clarity even though the same total numbers of particles, $N$, were
  used.  Note that the vertical range is expanded by a factor of 10
  for the points at the highest $N$ in the lower right panel.}
\label{fig.main}
\end{figure*}

\subsection{Measures of relaxation rate}
Figure~\ref{fig.main} summarizes the relaxation measurements from all
27 simulations.  The three rows show the three different mass models;
within each panel there are three different numbers of particles, and
for each case, the evolution was computed by three separate force
determination methods.  The measured rates from the heavy particles
are shown in red, while those from the light particles are marked in
green, and the uncertainties in the slopes are indicated by the error
bars, that are often too short to be visible.  The different force
determination methods used are distinguished by the different symbol
types as indicated in the figure caption, and offset horizontally from
each other for clarity even though the particle numbers are the same.
These conventions are the same in the left and right panels.

The left panels show the energy diffusion rate, defined as
\begin{equation}
\langle E_i(0)\rangle^{-2}\;{d\over dt}\langle[E_i(0)-E_i(t)]^2\rangle
\label{eq.edif}
\end{equation}
in units of $t_0^{-1}$.  The adopted values for $\langle
E_i(0)\rangle$, which are $-0.3$, $-0.45$, $-0.9$ for the Hernquist,
Plummer, and uniform sphere respectively, are the same for all nine
simulations with each model.  The straight dotted line has slope
$-0.5$ while the slope of the dashed line is $-1$; these lines are not
fits to the data and are for comparison only.  The right panels show
$Nd\langle E_i(t)\rangle/dt$, the factor $N$ for each sub-population
is included for clarity -- in reality, the slopes roughly decrease as
$1/N$.

There are many conclusions to be drawn from these data.

First, the differences between the energy diffusion rates (left
panels) for the heavy (red) and light (green) particles within one
simulation are generally small.  This is a further indication of the
dominance of coherent potential variations, which cause deflections
that are independent of particle mass, in driving this measure of
relaxation.

Second, the energy diffusion rate declines roughly as $N^{-1/2}$ in
the uniform sphere, whereas in the inhomogeneous models it declines
more or less with the expected $N^{-1}$ dependence for collisional
relaxation at fixed resolution (see \S1).  If two-body effects were
dominant in all cases, the variation with $N$ should be the same in
all three mass models.

Third, the rates of energy exchange shown in the right hand panels
generally have opposite signs, and vary roughly as $N^{-1}$, since
they are approximately constant when multiplied by $N$.  The values
from the highest $N$ experiments are quite uncertain, because the
decreasing variation in the mean as $N$ rises is masked by short-term
changes (see the right panel of Fig.~\ref{fig.typical}) -- \ie\ the
trends decrease into the noise.  This is particularly problematic for
the uniform sphere (bottom right panel), where the potential changes
associated with pulsations of the model dominate over the energy
exchange rate at the highest $N$, making an accurate measurement over
the time interval simulated impossible.

Fourth, the different methods used to compute the gravitational field
yield broadly similar behaviour in all three mass models, and the
variation of the rates with $N$ in both the left and right panels is
similar for each method in each of the different models. 

Fifth, the energy diffusion rate (left panels) is consistently lower,
but by a factor $<3$, for the SFP method (diamonds) than for the tree
code (crosses) and the spherical grid (circles) in the Hernquist and
Plummer models.  This systematic difference is also present in the
variations of $L$.  \citet{HO92} also reported a slightly lower energy
diffusion rate when they used a field method, although it was unclear
whether the more rapid diffusion in their tree code, for example,
resulted mainly from the mild disequilbrium of their initial model
caused by gravity softening.  However, the mass segregation rate
(right panels), which is a more direct consequence of two-body
scattering, is no less for SFP than for the other methods.

Sixth, relaxation in the uniform sphere probably is completely
dominated by the oscillatory modes excited by the shot noise in the
particle distribution.  The initial amplitudes of the modes should
scale as $N^{-1/2}$, and the decline in relaxation rate with
approximately this dependence in the uniform sphere is a direct
indication of their dominance in this case.  Collective modes are
present in all models, but the different $N$-dependence in the
inhomogeneous models is probably because modes are rapidly damped
in those cases.

Seventh, short-range gravity softening appears to have little effect
on the relaxation rate.  There is no explicit softening in SFP, and
the only smoothing on the spherical grid is linear interpolation in
radius.  Force softening is explicit in the tree code and implicit in
the C3D grid, where forces are slightly sharper $\epsilon_{\rm eff} =
0.036a$ while $\epsilon = 0.05a$ for the BHT code for the uniform
sphere.  Yet the relaxation rates scarcely differ between the various
codes.  This emphasizes the dominance of distant encounters, enhanced
by collective oscillations, in driving relaxation and that softening's
only value is to avoid large accelerations during close encounters
between particles, which would require short time steps to integrate
the motions accurately.

Once again, the quantities shown in Figure~\ref{fig.main} are averages
over all the particles, and these conclusions therefore apply to the
particles in the bulk of the model.  The small number of more tightly
bound particles have the largest energy changes, but by factors of
only a few, and therefore have little overall effect on the global
average.  Note also that their rate of energy change should be
reckoned on the time-scale of their shorter orbit periods.

\section{Discussion}
\subsection{Discrepancy with \citet{WK07a, WK07b}}
The finding stated in point four above is in strong disagreement with
that of \citet{WK07a, WK07b}, who argued that the relaxation
time-scale is orders of magnitude longer in field methods.
It should be noted that their analysis addressed the more limited
problem of resonant exchanges between particles and a perturbing bar
potential, not the more general relaxation studied here.  However, a
discrepancy remains to be explained since \citet{Sell08}, in a study
that addressed their specific prediction, also found that field
methods were not superior to grid methods.

Their calculation is highly technical but, in essence, they began by
separating large- from small-scale noise and calculated separate
contributions to the orbit deflections from both these ``components''
of the noise.  Rather surprisingly, they found that deflections due to
large-scale noise were much weaker than those from small-scale noise.
They then argued that direct codes possess noise on both scales while
field methods are affected only by large-scale noise, which led them
to conclude that relaxation should be far slower in field methods.
However, it is well-known that noise is present on all scales and,
since every decade of impact parameter contributes equally to the
Coulomb logarithm \citep{BT08}, there can be no clear distinction
between large- and small-scale noise.  This also implies that
relaxation from large-scale noise should be comparable to that from
small-scale noise.

\Ignore{
Typically, users of direct codes for collisionless problems adopt a
softening length that is ${\cal O}(1\%)$ of the effective radius of
the system, which provides at most only a slightly greater dynamic
range than is afforded by field methods when sufficient terms are
employed to follow significant mass rearrangement.  But even if we
accept the assumption by WK07 that the dynamic range in resolution in
most direct $N$-body simulations is orders of magnitude greater than
in field methods, its consequence should be slight because the dynamic
range affects the relaxation rate only through the Coulomb logarithm.}

\subsection{More on collective effects}
\citet{Vasi15} found good agreement between experimental relaxation
rates and theoretical predictions from Fokker-Planck diffusion
coefficients, suggesting that self-gravity of the collective modes does
not boost the relaxation rate significantly.  Furthermore, his
predicted rate for the Plummer sphere is in good agreement with the
value I find (Fig.~\ref{fig.main}, middle row, left panel).

However, such studies usually consider inhomogeneous models.  The
anomalous results presented here for the uniform sphere show that
relaxation {\em is} greatly enhanced by collective modes in this
unusual case where all orbits have the same period.  Note that
\citet{RT96} also found enhanced relaxation through collective effects
in their study of a system of low-mass particles in near Keplerian
motion about a central mass.  In their case, it was important that
particles with the same $L$ had the same period.

\subsection{Mild dependence on method}
The reason for the slightly lower diffusion rate in the field method
(point five above) is unclear.  Both the SFP and S3D methods employ an
expansion in surface harmonics up to $l_{\rm max}=8$ to capture any
angular variations, and the principal difference between them is
radial resolution: the SFP method employs 11 radial functions ($0 \leq
n \leq 10$), while the S3D uses a grid of $\sim100$ radial shells.

It therefore seemed that the marginally higher energy diffusion rate
when the S3D grid was used was because that grid could support more
oscillatory modes.  However, increasing the number of radial functions
used in the SFP method to 51, in order to enable the field method also
to support more collective modes, led to little change in either
measure of the relaxation rate, confounding this theory!  Additional
experiments with $l_{\rm max}=4$ and $l_{\rm max}=16$ caused very
slight changes in the measured relaxation rates in the expected sense,
but the systematic discrepancy between the two methods persisted.

While the puzzle remains, the negative outcome of these tests is
further evidence of the unimportance of self-gravity for collective
oscillations in inhomogeneous models.

\section{Conclusions}
The main result here confirms that previously found by \citet{HB90}
and by \citet{HO92} that the rate of relaxation in $N$-body
simulations that aspire to be collisionless is very largely
independent of the method used to compute the gravitational field.
This is especially true when the relaxation rate is assayed as the
rate of energy exchange between particles of different masses -- see
the right hand panels of Fig.~\ref{fig.main}.  This result is
physically reasonable, since relaxation is dominated by distant
encounters and any method that correctly yields the field from distant
particles must faithfully include their stochastic contributions.

The rate of relaxation arises from at least two distinguishable
sources: a slow diffusion of the integrals caused by coherent
potential oscillations of the system that are largely independent of
particle mass, and mass segregation that is more directly caused by
two-body encounters.  The good agreement (\S5.2) between Fokker-Planck
diffusion and $N$-body experiment in the inhomogeneous models
indicates that self-gravity of the collective modes adds little to the
relaxation rate.  The rate of energy exchange between particles of
differing masses becomes more difficult to measure as $N$ rises
because mean energies scatter with potential variations that scale as
$N^{-1/2}$, while the trends in the mean energies of the different
mass species diverge as $N^{-1}$.

The different $N$-dependence in the uniform sphere (bottom left panel
of Fig.~\ref{fig.main}) is clear evidence that collective modes {\em do}
dominate the relaxation in this special case.  This model differs from
the other two by having a harmonic potential throughout in which the
orbit frequencies of all particles are the same; this difference
permits undamped collective oscillations, whereas collective modes are
damped in the other cases.

The slightly lower energy diffusion rates that result from use of the
field method (diamonds in top left and middle left panels of
Fig.~\ref{fig.main}) is bought at a high price.  The leading term in
the basis used for both cases was a perfect match to the equilibrium
model, and the same basis would be less suited were the density to
evolve, or were it used for any other model, requiring more terms to
yield the correct total potential.  Thus field methods lack the
versatility to follow arbitrary changes to the distribution of mass
within a model unless the expansion is taken to higher order, and
their slightly better relaxation rate can be achieved in any of the
more general methods simply by employing a few times more particles.
A real advantage of field methods, featured by \citet{WK07a, WK07b},
is that they offer an elegant comparison of simulations with
perturbation theory when computing first order changes to an
equilibrium model that could be unstable or externally perturbed.

The energy diffusion time-scale, the inverse of the rate defined in
eq.~(\ref{eq.edif}) and plotted in Fig.~\ref{fig.main}, is $\ga 10^5$
dynamical times (\ie\ $10^{12}$~yr for the suggested scaling) for only
$N=4\times 10^5$ particles in the inhomogeneous models (it is shorter
for the exotic case of the uniform sphere).  By this measure,
relaxation times in simulations of inhomogeneous 3D models using any
valid code are already significantly longer than the ages of the
galaxies being simulated, and relaxation considerations alone do not
require much larger numbers of particles.

\section*{Acknowledgments}
The author wishes to thank Tad Pryor for many insightful conversations
and the referee for a perceptive report.  I also thank Eugene Vasiliev
for helpful comments on an earlier draft and Martin Weinberg for an
extensive e-mail correspondence.  This work was supported by NSF grant
AST/1108977.

\label{lastpage}


\begin{thebibliography}{}

\def\aap{A\&A}
\def\aj{AJ}
\def\apj{ApJ}
\def\apjl{ApJL}
\def\apjs{ApJS}
\def\apss{Ap.\ Sp.\ Sci.}
\def\araa{ARAA}
\def\jcop{J. Comp.\ Phys.}
\def\mnras{MNRAS}
\def\newa{New. Astron.}
\def\PhD{PhD.\ thesis}
\def\nat{Nature}
\def\pf{{\it Phys.\ Fluids}}
\def\PhD{{\it PhD thesis}}
\def\phya{{\it Physica\/} A}
\def\rpp{Rep.\ Prog.\ Phys.}
\def\sovast{{\it Sov.\ Astron.}}

\bibitem[\protect\citeauthoryear{Barnes \& Hut}{1986}]{BH86}
Barnes, J. \& Hut, P. 1986, \nat, {\bf 324}, 446


\bibitem[\protect\citeauthoryear{Binney \& Tremaine}{2008}]{BT08}
Binney, J. \& Tremaine, S. 2008, {\it Galactic Dynamics\/} (2nd ed.; Princeton: Princeton University Press)

\bibitem[\protect\citeauthoryear{Chandrasekhar}{1941}]{Chan41}
Chandrasekhar, S. 1941, \apj, {\bf 94}, 511


\bibitem[\protect\citeauthoryear{Clutton-Brock}{1972}]{CB72}
Clutton-Brock, M. 1972, \apss, {\bf 17}, 292

\bibitem[\protect\citeauthoryear{Debattista \& Sellwood}{2000}]{DS00}
Debattista, V. P. \& Sellwood, J. A. 2000, \apj, {\bf 543}, 704


\bibitem[\protect\citeauthoryear{H\'enon}{1973}]{Heno73}
H\'enon, M. 1973, in {\it Dynamical Structure and Evolution of Stellar Systems}, ed.\ L. Martinet \& M. Mayor (Sauverny: Geneva Observatory) p.~182

\bibitem[\protect\citeauthoryear{Hernquist}{1990}]{Hern90}
Hernquist, L. 1990, \apj, {\bf 356}, 359

\bibitem[\protect\citeauthoryear{Hernquist \& Barnes}{1990}]{HB90}
Hernquist, L. \& Barnes, J. E. 1990, \apj, {\bf 349}, 562

\bibitem[\protect\citeauthoryear{Hernquist \& Ostriker}{1992}]{HO92}
Hernquist, L. \& Ostriker, J. P. 1992, \apj, {\bf 386}, 375


\bibitem[\protect\citeauthoryear{Hockney \& Eastwood}{1981}]{HE81}
Hockney, R. W. \& Eastwood, J. W. 1981, {\it Computer Simulation Using Particles
\/}, New York:McGraw Hill


\bibitem[\protect\citeauthoryear{Holley-Bockelmann, Weinberg \& Katz}{2005}]{HBWK05}
Holley-Bockelmann, K., Weinberg, M. \& Katz, N. 2005, \mnras, {\bf 363}, 991

\bibitem[\protect\citeauthoryear{James}{1977}]{Jame77}
James, R. A. 1977, \jcop, {\bf 25}, 71 




\bibitem[\protect\citeauthoryear{McGlynn}{1984}]{McGl84}
McGlynn, T. A. 1984, \apj, {\bf 281}, 13

\bibitem[\protect\citeauthoryear{Merritt}{2013}]{Merr13}
Merritt, D. 2013, {\it Dynamics and Evolution of Galactic Nuclei\/} (Princeton: Princeton University Press)

\bibitem[\protect\citeauthoryear{Monaghan}{1992}]{Mona92}
Monaghan, J. J. 1992, \araa, {\bf 30}, 543

\bibitem[\protect\citeauthoryear{Plummer}{1911}]{Plum11}
Plummer, H. C. 1911, \mnras, {\bf 71}, 460

\bibitem[\protect\citeauthoryear{Polyachenko \& Shukhman}{1979}]{PS79}
Polyachenko, V. L. \& Shukhman, I. G. 1979, Astron.\ Zh.\ {\bf 56}, 724; English translation: 1981, \sovast, {\bf 25}, 533

\bibitem[\protect\citeauthoryear{Rauch \& Tremaine}{1996}]{RT96}
Rauch, K. P. \& Tremaine, S. 1996, \newa, {\bf 1}, 149

\bibitem[\protect\citeauthoryear{Rostoker \& Rosenbluth}{1960}]{RR60}
Rostoker, N. \& Rosenbluth, M. N. 1960, \pf, {\bf 3}, 1



\bibitem[\protect\citeauthoryear{Sellwood}{2003}]{Sell03}
Sellwood, J. A. 2003, \apj, {\bf 587}, 638

\bibitem[\protect\citeauthoryear{Sellwood}{2008}]{Sell08}
Sellwood, J. A. 2008, \apj, {\bf 679}, 379



\bibitem[\protect\citeauthoryear{Sellwood}{2013}]{Sell13}
Sellwood, J. A. 2013, \apjl, {\bf 769}, L24

\bibitem[\protect\citeauthoryear{Sellwood}{2014}]{Sell14}
Sellwood, J. A. 2014, arXiv:1406.6606 (on-line manual: \hfil\break {\tt http://www.physics.rutgers.edu/$\sim$sellwood/manual.pdf})

\bibitem[\protect\citeauthoryear{Sellwood \& Merritt}{1994}]{SM94}
Sellwood, J. A. \& Merritt, D. 1994, \apj, {\bf 425}, 530


\bibitem[\protect\citeauthoryear{Spitzer}{1987}]{Spit87}
Spitzer, L. 1987, {\it Dynamical Evolution of Globular Clusters\/} (Princeton: Princeton University Press)

\bibitem[\protect\citeauthoryear{Springel}{2005}]{Spri05}
Springel, V. 2005, \mnras, {\bf 364}, 1105





\bibitem[\protect\citeauthoryear{Vasiliev}{2015}]{Vasi15}
Vasiliev, E. 2015, \mnras, {\bf 446}, 3150

\bibitem[\protect\citeauthoryear{Weinberg}{1998}]{Wein98}
Weinberg, M. D. 1998, \mnras, {\bf 297}, 101

\bibitem[\protect\citeauthoryear{Weinberg}{1999}]{Wein99}
Weinberg, M. D. 1999, \aj, {\bf 117}, 629

\bibitem[\protect\citeauthoryear{Weinberg \& Katz}{2007a}]{WK07a}
Weinberg, M. D. \& Katz, N. 2007a, \mnras, {\bf 375}, 425

\bibitem[\protect\citeauthoryear{Weinberg \& Katz}{2007b}]{WK07b}
Weinberg, M. D. \& Katz, N. 2007b, \mnras, {\bf 375}, 460



\end{thebibliography}
\end{document}